\documentclass[apjl]{emulateapj}

\usepackage{graphicx}
\usepackage{amsmath,amssymb}
\usepackage{epstopdf}

\usepackage {pstricks}

\begin{document}

\title{Key Issues Review: Numerical studies of turbulence in stars}

\author{W. David Arnett\altaffilmark{1}
}
\author{Casey Meakin\altaffilmark{1} }

\altaffiltext{1}{Steward Observatory, University of Arizona, 
933 N. Cherry Avenue, Tucson AZ 85721}




{\bf DRAFT FROM \today}

\begin{abstract}
The numerical simulation of turbulence in stars has led to a rich set of possibilities regarding stellar pulsations, asteroseismology, thermonuclear yields, and formation of neutron stars and black holes. The breaking of symmetry by turbulent flow grows in amplitude as collapse is approached, which insures that the conditions at the onset of collapse are not spherical. This lack of spherical symmetry has important implications for the mechanism of explosion and ejected nucleosynthesis products. 
Numerical resolution of several different types of three--dimensional (3D) stellar simulations are compared; it is suggested that core collapse simulations may be under-resolved.
 New physical effects which appear in 3D are summarized.
 Connections between simulations of progenitor explosion and observations of supernova remnants (SNR) are discussed.
 Present treatment of boundaries, for mixing regions during He--burning, requires revision.
\end{abstract}

\section{Introduction}\label{Sintro}
Most of the history of the theory of stellar evolution is based on the approximation that stars have spherical symmetry, and are quasi--static objects. This flies in the face of observations of the surface of the Sun, as time--lapse videos\footnote{e.g., https://svs.gsfc.nasa.gov/cgi-bin/details.cgi?aid=3412 }  of sunspots, eruptions and granulation show, so that the approximation must be stated more carefully as {\em spherically symmetric on average.} The Sun is a gravitationally--bound thermonuclear reactor, made of plasma which exhibits turbulent convection in its outer layers. {\em Spherical symmetry on average} arises from the cancellation of effects, from the combined action of many chaotic fluctuations such as those which the solar videos capture. The fortunate approximations of spherical symmetry and quasi-static behavior allow the reduction of a three-dimensional (3D) time--dependent problem to a one dimensional (1D) quasi--static one, which is amenable to numerical simulation by computer as fifty years or so of progress have shown.

Direct simulation of turbulent convection in stars is now feasible, and permits a theoretical study of the solutions to the 3D, time--dependent equations of fluid  flow in stars, and consequently a critical re-examination of the spherical, quasi-static approximation for stellar evolution.

\section{Methods}\label{Smethods}

\begin{deluxetable*}{lllll}
\tablewidth{280pt}
\tablecaption{Some turbulence simulations with RANS analysis \label{table1}}
\tabletypesize{\small}

\tablehead{  \colhead{reference} &\colhead{type} & \colhead{code} 
& \colhead{grid size}   & \colhead{name}
}
\startdata
 \cite{ma07b} & OB$^a$ & PROMPI &  $400\times 100^2 $ & lo-res \\
\\
 \cite{viallet2013}  & OB & PROMPI & $786 \times 512^2$  & hi-res\\
   & OB & PROMPI & $384 \times 256^2$ & med-res\\
    & OB & PROMPI & $192 \times 128^2$ & lo-res\\
\\
   & RG$^b$ & MUSIC & $ 432 \times 256^2 $ & m-med-res\\
     & RG & MUSIC & $ 216 \times 128^2 $ & m-lo-res\\
\\  
   S. Campbell$^c$ and & OB & PROMPI & $1536 \times 1024^2 $ & ultra-hi-res \\
      \cite{321D} & \\
\enddata

\tablenotetext{a}{Oxygen burning shell in pre--collapse star of 23~$M_\odot$.}
\tablenotetext{b}{Red giant envelope convection, \cite{viallet2011}.}
\tablenotetext{c}{Paper in preparation.}

\end{deluxetable*}

\placetable{1}

Table~\ref{table1} summarizes the simulations \citep{ma07b,viallet2013,321D} upon which this discussion is based. In 3D the turbulent cascade moves from large to small scales, so the implicit large--eddy simulation
approach (ILES) of \cite{boris} maximizes effective use of computational resources. To attain higher resolution a
``box--in--star'' grid configuration is used, so that only a part of the star is simulated on the grid. 
This gives a maximal numerical Reynolds number, and truly turbulent behavior using presently available computers, with the drawback that the limited size of the box, being smaller than the star, does not support the lowest order modes of motion. 
A resolution study (a range of 8 in linear resolution, or $8^4=4096$ in computational resources) suggests that the turbulent cascade and the convective boundaries may be adequately captured \citep{321D}.
These simulations use realistic microphysics for the equation of state (EOS), opacity, nuclear reaction network (25 nuclei), and neutrino emission. The initial model was mapped from a well--resolved 1D stellar evolutionary model onto the 3D computational grid.  The residuals from conservation laws are monitored to test both the use of the ILES approximation, and the zoning in boundary layers \citep{viallet2013,321D}.

A number of other efforts have contributed to our improved understanding of 3D stellar convection.
\cite{herwig,herwig11} have conducted 3D simulations of similar resolution and character (on proton ingestion during He flash convection), using a ``star--in--box'' approach (the entire star is enclosed by the computational grid, allowing even the lowest order modes to be represented)\footnote{This is at the cost of requiring correspondingly greater computational resources; the grids were cubes of $768^3$, $1152^3$ and $1536^3 $ cells. An improved numerical scheme was used to provide a higher effective resolution. The $1536^3$ run required four days on Blue Waters at  the National Center for Supercomputing Applications (NCSA). Also see Table~\ref{table1} for comparison.}. See earlier work in \cite{pw94,porter99,pw00}. \cite{simon16a} show a ``box--in--star'' simulations of the upper boundary of an $8\,\rm M_\odot$ star, which is a related ingestion problem.

The earliest successes of 3D radiative hydrodynamics were the pioneering simulations of the solar atmosphere by Nordlund and Stein, which reproduced the observed shapes of spectral lines without adjustable parameters. See \cite{aake85,nordstein,nsa,sn89,sn98,magic2013}. These simulations were in the ``box--in--star'' mode, and treated magnetic fields on  the grid in the MHD approximation, but contained no composition gradients or nuclear burning.
See \cite{fls96,lfs99,co5bold,ludwig-rg} for further work on 3D stellar atmospheres.

Juri Toomre and collaborators had a different focus, on rotation and magnetic fields in the ``star--in--box'' mode,  developing the MHD anelastic code ASH \citep{mltz84,hurl84,hurl86,catt91,bt02,brun-miesch-toomre,bp09,bmj_ash,miesch,mbdt08}. See also the discussion concerning energy conservation\footnote{Residuals from conservation laws are useful diagnostics for resolution and consistency \citep{viallet2013}, which \cite{mbdt08} also show to be an issue.} in \cite{bvz12,vlbwz13}. These simulations used a simplified atmospheric boundary, and contained no composition gradients or nuclear burning, but considered rotation and magnetic fields in detail.

 Adaptive mesh refinement (AMR) is an attempt to attain better resolution with limited computational resources; this has been implemented in several stellar astrophysics codes: the compressible hydrodynamics code FLASH \citep{flash}, the low--Mach--number hydrodynamics code MAESTRO \citep{maestro10}, and the compressible hydrodynamics code CASTRO \citep{castro}. The basic idea is to use fine zoning in regions where it is needed and coarser resolution elsewhere. For this to be effective there must be regions on the grid in which little is happening, so the degree of effectivenes of AMR may depend upon the problem being tackled. For example, turbulence expands to fill the volume allowed, and be a broadly distributed, while in contrast, nuclear burning is highly sensitive to temperature, and may be localized.
 
Extensive theoretical, numerical and experimental information
from the geosciences has been analysed by \cite{can11a,can11b,can11c,can11d,can11e}, for application to the problem of stellar convection and mixing.

Given the variety in emphasis, resources, and interests, these 3D simulations are strikingly consistent, which encourages this attempt to explore their implications.

\section{Results}\label{Sresults}

\begin{figure}[h]
\figurenum{1}

\includegraphics[angle=0,scale=0.30]{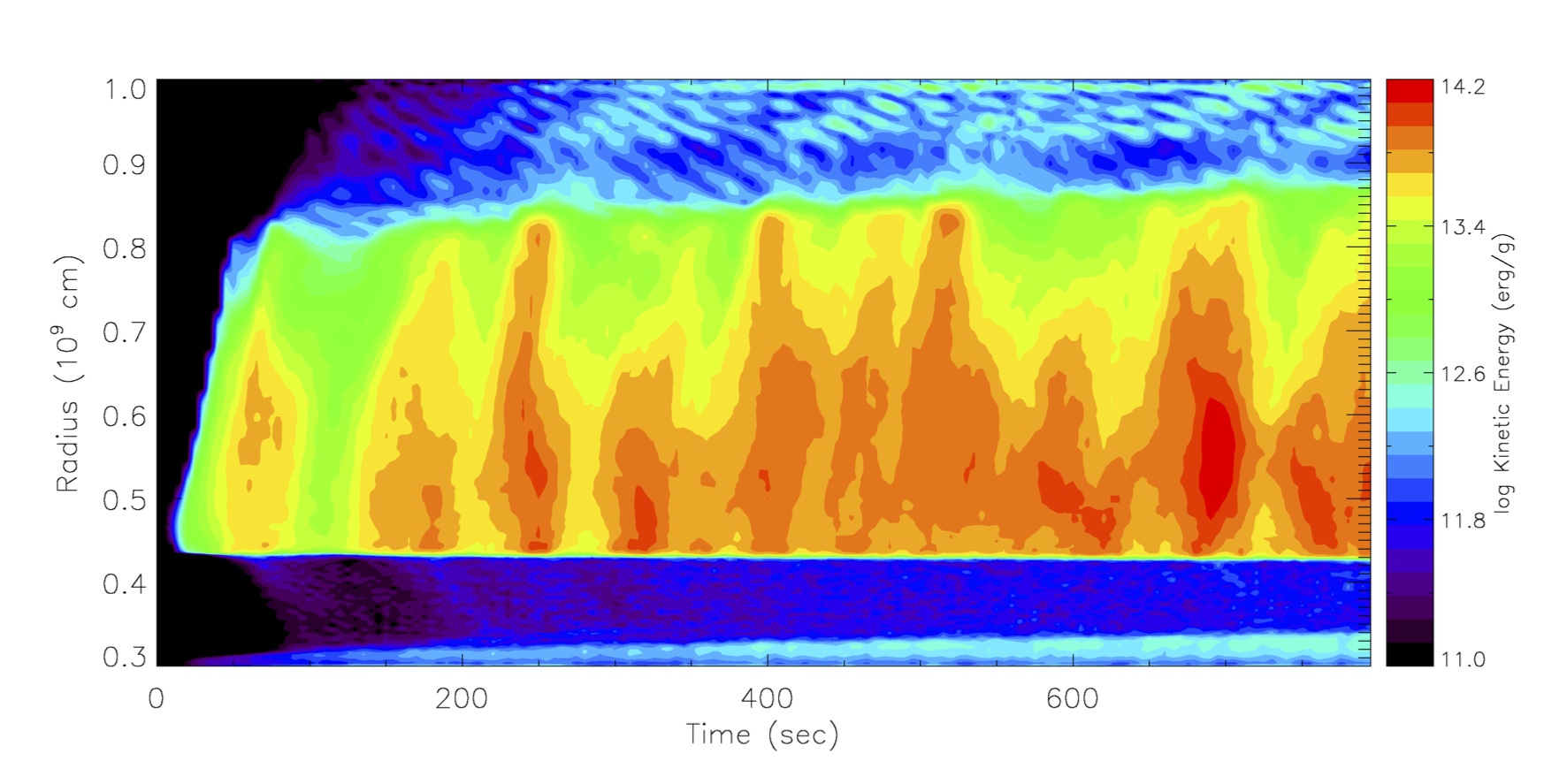}

  \caption{The time evolution of the radial profile of contours of turbulent specific kinetic energy
    is shown for the 3D oxygen shell burning model, 
     which illustrates the
    intermittent nature of the convective motions, even in a quasi--steady state.  
    The upwelling chimney--like
    features in the convective region are seen to excite internal wave
    trains in the stable layers, which can be seen to propagate away from the
    boundaries of the convection zones. Entrainment erodes both the top and bottom boundaries of convection,    
    so the convective region grows.  {\it After \cite{ma07b}.}
    \label{ob3d-tke}}
\end{figure}

Qualitatively new and previously overlooked behavior appears in the simulations indicated in Table~\ref{table1} and referred to in \S\ref{Smethods} above.

\subsection{Dynamics}
Stellar turbulent convection is a strikingly dynamic process \citep{ma07b}. This is indicated in Fig.~\ref{ob3d-tke}, which shows the dynamic onset of convection from a carefully zoned 1D initial model mapped to a 3D grid. Because of the inadequacy of present algorithms for 1D turbulent convection, there is always an initial transient in which convective flow is established.  This is followed by 
bursts of turbulent kinetic energy, recurring on a time interval similar to the time needed to transit the convective layer. These bursts encounter stable layers, and drive wave motions (predominantly gravity modes) from both boundaries, most obviously from the top. The boundaries are eroded by entrainment. Both the rate of entrainment, and the amplitude of the waves, are affected by the stiffness of the stable layer (the value of the Brunt-V\"ais\"al\"a frequency), and are not some universal constant.
The behavior is also modified by radiative effects, as shown by comparison of the oxygen--burning shell (OB) and the lower boundary of the surface convection zone of a  red giant (RG) in  \cite{viallet2013}, which have significantly different values of the P\`eclet number \citep{viallet2015}.

\subsection{Reynolds--averaged Navier-Stokes equations}
At present the highest resolution grids have $\sim 10^9$ grid points, which for a 25-nucleus network implies a storage requirement of $\sim 3 \times 10^{11}$ individual numbers per timestep. The challenge is to abstract the essential features from this flood of data. The Reynolds-averaged Navier--Stokes (RANS) equations provide a systematic way to condense the data and to provide insight for the development of approximate equations to describe it  \citep{miro2014}. Two powerful features of this formulation are that it allows a quantitative evaluation of the sub--grid dissipation, and it explicitly allows the identification of errors due to inadequate resolution \citep{viallet2013}. The latter is particularly useful in assessing the numerical quality of simulations of boundaries between turbulent and non-turbulent regions.
 
\subsection{Kolmogorov cascade}
There is a balance, on average, between buoyancy and turbulent dissipation \citep{amy09vel,am11turb}.
Buoyancy occurs as a radial acceleration at the largest scales. At the large Reynolds numbers typical of stars, the motion is unstable and breaks--up into smaller scale motion, becoming isotropic and losing its 1D character.  This continues until the Kolmogorov scale is reached, at which point dissipative processes can damp the motion. For these ILES simulations this damping occurs at the sub--grid scale. With sufficient resolution (e.g., see Table~\ref{table1}, \cite{herwig,nonaka-snia}), the turbulent cascade is reasonably well approximated.

\subsection{Lorenz convective roll}
The largest scale (the {\em integral} scale) behaves like the convective roll of \cite{lorenz}, which is a classic example of a dynamic system with a strange attractor and chaotic behavior \citep{manneville}. A combination of the Lorenz model and the Kolmogorov cascade give a simple but qualitatively consistent representation of the 3D simulation \citep{amy09vel,am11turb}. The steady--state limit of this combination gives the {\em steady vortex} model, from which the mixing--length theory (MLT) cubic equation of \cite{bv58} can be derived \citep{nathan2014}, a process which illustrates the strength and limitations of MLT\footnote{MLT is the standard theory of convection now used in stellar evolution.}. 

\subsection{Fluctuations and Rayleigh--Taylor Instability}
If a steady--state constraint is not imposed, the balance between buoyant driving and turbulent damping is not satisfied instantaneously, but only on average. Because driving and damping occur at widely different length scales, they are not synchronized in time, causing the intermittent pulse behavior seen in Fig.~\ref{ob3d-tke}.
If the driving is sufficiently strong, an explosion  may develop instead of steady--state convection, with the rising plumes becoming Rayleigh--Taylor (RT) unstable and, depending upon the structure, turbulent \citep{swisher}. Expansion causes a ``freeze--out'' of the motions \citep{abarzhi10} and of the compositional structure, tending toward a spherical explosion (a Hubble flow).

\subsection{Turbulent Kinetic Energy Flux}
Convective regions possess a basic up--down asymmetry, which increases with stratification. In a hydrostatic stratification, rising material expands because it encounters decreasing pressure, and visa versa. To conserve mass, rising plumes will tend to be broader and slower than descending plumes. This difference requires a net downward flux of turbulent kinetic energy (TKE) \citep{ma10apss}. This is pronounced in convective atmospheres, which are strongly stratified, containing many pressure scale heights. Net flux of turbulent kinetic energy is assumed to be zero in MLT,
 but is nonzero and a prominent feature of 3D simulations of stellar atmospheres \citep{ns95,nsa}.


\subsection{Boundary layers}
A key feature of sufficiently resolved 3D simulations \citep{herwig,321D} is the development of thin layers which separate regions of potential flow (non-turbulent) from those of solenoidal flow (turbulent).
Because MLT is a local theory, it has no ability to describe such boundaries. By convention this gap is addressed by linear stability analysis, resulting in use of the criteria of Schwarzschild (no composition gradient) and Ledoux to define boundaries of mixing.  In the geosciences a more sophisticated analysis leads to various versions of the Richardson criterion (see \cite{turner}, \S10.2.3), which includes both the degree of convective vigor and the stiffness of the stable region. The Richardson criterion is a ratio of the potential energy required to mix a stably stratified layer, divided by the turbulent kinetic energy available to do so.
This preference of a Richardson criterion for mixing is supported by the simulations (\S3.5, \cite{321D}). 
The Ledoux criterion only includes the effects of a composition gradient, but not convective vigor, and so underestimates mixing\footnote{Both thermonuclear burning by fusion and microscopic diffusion tend to develop stable composition gradients in stars.}. This may be why the Schwarzschild criterion is preferred over the Ledoux. The Schwarzschild criterion ignores the inhibiting composition gradient term, is simpler, and better agrees with astronomical observations. 
Accurate estimates of {\em both} the composition gradient and the convective speed are needed to define the true mixing boundary. 
 
 \subsection{Composition  currents}
 The mixing of composition in the simulations is advective \citep{321D} rather than diffusion--like \citep{ppe72}, and transport is dominated by the large eddies. Rising and descending flows have differing composition, giving rise to {\em composition currents}. Mathematically this reduces the highest order spatial derivative by one: the composition flux is propotional to the composition, not its gradient. In practice this may result in a qualitatively similar difference operator in 1D \citep{am11real}, but caution is suggested in use of the Eggleton algorithm. While it is robust and useful for rough surveys, it may be inappropriate for more precise use. In particular, it may not be the best foundation for introducing rotational effects \citep{maeder}. A better approach might involve using the integral--scale flows and rotation directly \citep{balbus09,mbdt08}.

\section{Discussion}\label{Sdiscuss}

\subsection{Increasing asymmetry approaching  collapse}\label{s-progen}
As indicated in \S\ref{Sintro}, the validity of the spherically-symmetric approximation for the Sun depends upon a cancellation of effects from many turbulent cells acting incoherently.
Neutrino emission reduces the thermal time scale relative to the hydrodynamic time scale (the sound travel time), so that it becomes computationally easier\footnote{Solar simulations may use ``box-in-star'' to avoid the slow thermal relaxation time of the deep layers \citep{nsa}, or simplify the outer boundary condition and use ``sound-proofing'' methods \citep{bvz12} to allow larger time steps \citep{brun-miesch-toomre}.} to directly simulate the late stages of stellar evolution \citep{wda96}.

As collapse is approached, temperature rises. Neutrino cooling increases, balanced by increased heating from nuclear burning, which gives increased turbulent velocity \citep{wda96}. 
The increased vigor of convection causes increased fluctuation amplitudes for density as well as tangential velocity. 
The core is the least stratified part of the star, so that the convective cells are larger and fewer, giving a less smooth cancellation of chaotic behavior.
As collapse is approached, the equation of state (EOS) is softer (the adiabatic exponent $\Gamma \rightarrow 4 / 3$), so that radial restoring forces decrease, giving larger amplitude fluctuations.
  
This tendency for dynamic behavior was already indicated in 2D simulations, \citep{ba94,ba97a,ba97b,ba98,aa00,ma06,am11real}.
At present, pre-collapse progenitor models have spherical symmetry, and no fluctuations or tangential velocities
 (e.g., \cite{ww95,whw02}). The simulations, both 2D and 3D, show that these progenitor models represent an extreme and non--physical limiting case. 
 
The problem of inertial confinement fusion (ICF)
 is similar to core collapse in that a compression by a large factor ($\sim100$) is involved \citep{rem99,rem00}. Experiments show the extreme importance of slight asymmetries in the initial state prior to compression; these can modify even the qualitative nature of the flow \citep{drake,lindl}, a feature shared with core collapse \citep{cott13}.
  
\subsection{ Precollapse dynamics}\label{s-predyn}
In addition to changes in shape, 3D simulations also show significant additional changes in time, which also can affect the evolution up to and into collapse.
Waves are generated at the convective boundaries (Fig.~\ref{ob3d-tke}), which give rise to entrainment  and mass loss \citep{ma07b}. For the precollapse stages, these waves may not have time to travel to the stellar surface, and they may dissipate along the way.
 The mass loss may be extensive \citep{quat_shiode,shiode_quat}, and the dynamics may even lead to eruptions prior to collapse \citep{am11real}, as suggested by observations of SNIIn \citep{nathan2014}.
 In order to get ``Type I'' light curves for SNIbc supernovae, significant mass loss must occur prior to collapse
 \citep{wda82}; wave--driven mass--loss and eruptions may be alternatives to binary stripping \citep{wda80}.
 
 The 3D morphology of silicon--burning \citep{couch2015} changes from that of oxygen--burning, because the time scale for nuclear burning approaches the turnover time scale for convective flow. The further evolution of Si--burning to nuclear statistical equilibrium (NSE)  in a convection zone corresponds to a system having both mechanical and thermodynamic degrees of freedom (see \cite{llsp}, \S123), an unexplored\footnote{Previously the mechanical degrees of freedom were ignored during the approach to equilibrium; see \cite{clay83}.} theoretical  aspect of the approach to collapse indicated by 3D simulations.
 
 \subsection{Implications for core--collapse simulations}\label{s-collapse}
 
 Core collapse of a massive star leads to the formation of a neutron star or a black hole, but the details of this process have presented continuing difficulty\footnote{See the tortuous development in \cite{wda67,wda77,wda77b,wda80,wda96} for example.}.
\cite{janka16} have recently reviewed the physics of core--collapse in 3D.
Core collapse simulations add the calculation of neutrino spectrum transport to the computational challenges already in 3D simulations of \cite{321D,herwig}, and so cannot afford equal resolution. 
In addition, core collapse simulations require a star--in--box approach (see \S\ref{Smethods}); a full $4\pi$ geometry is necessary.
The Oak Ridge group used the CHIMERA code with a $540\times 180^2$ grid, and the Garching group used the PROMETHEUS-VERTEX code with a $600 \times 90 \times 180$ grid. If these were box-in-star simulations, they would both correspond to the low--resolution examples in Table~\ref{table1}. However, as star-in-box simulations,  they correspond to a resolution still coarser by roughly a factor $\sim 4 \times 8 =32$, as only $1/32$ of the stellar volume needed to be simulated in the box--in--star case, with a comparable number of zones. The star--in--box simulation of \cite{herwig} used a grid of $1536^3$ points {\em and} a refined numerical algorithm to improve resolution (see their appendix).

For comparison, \cite{sn98} used grids of $125^2 \times 82$ and $ 253^2 \times 163$ for satifactory resolution of the solar atmosphere, but in the box--in--star mode; they  computed a $ 6^2 \times 3 $ Mm$^3$ box ($1.88\times10^{26} \rm \ cm^3$), or $7.5 \times 10^{-8}$ of the solar volume. Clearly a star--in--box simulation, which is necessary to represent core collapse, would require additional zones to maintain resolution. This extreme requirement is softened, but not removed, by the fact that the solar opacity is more temperature dependent that the neutrino opacities (giving stronger gradients), and the solar density gradient is steeper (requiring more convective cells) for the solar case.

A comparison of zoning in the ASH code is more complex.  \cite{mbdt08} quote a value is $ 257 \times 1024 \times 2048$ for a star--in--box simulation, and state that ``previous simulations did not have sufficient resolution to capture such dynamics'', which is consistent with the general trend discussed above.
 
 With their lower demand on computational resources, 2D simulations might seem tempting, but they are inadequate for turbulent simulations due to the different cascade direction relative to 3D. The flows are simply different; see Fig.~4 in \cite{ma07b}. Simulations in 2D is more useful for debugging code (and ideas) than describing stars.

The 3D simulations in Table~\ref{table1} suggest that bulk features of turbulence are reproduced even in the low--resolution cases, but that more resolution is needed for boundaries. Are the problem areas discussed by \cite{janka16} more akin to bulk features or to boundaries? The
gain--region is sensitive to differences in tangential flow \citep{cott13}, a feature itself affected by resolution.
The standing accretion shock instability (SASI, \cite{sasi}) seems more global, and might be captured with even the coarse resolution now used. 
The lepton--number emission asymmmetry (LESA, \cite{lesa}) may be similar to convection currents seen in O-burning and proton ingestion simulations  at higher resolution (see \S\ref{Sresults}). Comparison of Fig.~4 in \cite{janka16} to Fig.~2 in \cite{herwig} and Fig.~1 in \cite{mbdt08}, all star--in--box simulations, suggests that higher resolution may be needed for the collapse simulations.


 The 3D simulations of pre--collapse stages also have implications for the collapse process.
They suggest that the boundaries may become steeper during the neutrino--cooled stages, as collapse is approached. This would affect core size and density stratification in the progenitor.
It may decrease the rate of accretion on the newly-formed neutron star; this would reduce the photodisentegration losses which kill the original bounce shock and oppose shock rebirth. 
A reduced accretion rate might also result from pre--collapse eruption or extensive mass loss.

 It should be stressed that the integral scale turbulence carries the most momentum and kinetic energy, does the most transport, and is the least isotropic. Consequently the isotropy of collapse, or lack thereof, is an important feature.
 
\subsection{Implications for thermonuclear supernovae}\label{s-thermonuclear}

 Carbon--burning under conditions of high electron degeneracy begins with a stage of convective instability (``simmering") which temporarily delays the thermonuclear runaway \citep{wda68,wda69b}. This insures that the runaway begins in a fully 3D and chaotic manner. 
 The time dependent convection algorithm originally used was similar to that subsequently derived from 3D simulations \citep{321D}.
 Under--resolved simulations would tend toward laminar flow (a dipole velocity field), but finer zoning results in chaotic, turbulent flow. 
 
 
The first simulations of the delayed detonation model of thermonuclear supernovae,  with detailed nucleosynthesis and relatively high 3D resolution, were  done by \cite{seitenzahl2013}, using a $512^3$ grid and the LEAFS code. They did not model the
 ignition phase. They treated the ``unresolved sub-grid acceleration of flame due to turbulence'' with a sub-grid model, and did nucleosynthesis by tracer particles.
  \cite{fink2014} did pure deflagration models with the same code and  grid. 
 
The simmering phase was investigated by \cite{nonaka-snia} and \cite{zingale-maestro11} using the low-Mach code MAESTRO,   with grids of $384^3$ and $576^3$; MAESTRO allowed time steps which were up to a factor of 18 larger\footnote{While useful for this problem, this speed--up is  unfortunately far too small to be of value for most stellar evolutionary stages.} than a compressible code. After the burning developed, this simulation was then continued with the compressible hydrodynamics code
CASTRO,  \citep{malone-castro-14}; using AMR the effective resolution was quoted as $1152^3$.

Extension of simulations to investigation of the
boundary between core collapse and thermonuclear explosion of ONe cores  \citep{sjones-ONe},
and to the white dwarf merger channel for SNIa and merger in general and
common envelope evolution \citep{ohlmann2016comenv} is beginning to be made. The resolution issue seems particularly acute for mergers because of the steep gradient at the surface  of each star (a boundary problem). 
 
\subsection{Young SN remnants}\label{s-ysnr}

There is a deep connection between convection and the Rayleigh-Tayor instability (RTI).
The analysis of the RTI leads to differential equations \citep{abarzhi10,swisher} which are similar to those
resulting from analysis of the 3D simulations of turbulence \citep{321D}. Roughly speaking, convection starts from a plume which breaks up into the turbulent cascade, and being contained in a given volume, repeats the process (Fig.~\ref{ob3d-tke}).
The RTI begins in a similar way, but is uncontained, so that expansion converts internal energy into kinetic energy by $PdV$ work. In this sense, the RTI is a process of {\em convective freeze-out}. The RTI converts the initial compositional structure of a supernova progenitor into a supernova remnant, so that the composition, distribution, and velocity of the filaments contain information about both the progenitor and the explosion. This is why young supernova remnants can be poorly mixed
\citep{hughes_casA00}.

This became clear with Cas A \citep{kc-casA,ck-casA}, which showed the previously predicted compositional signature of incomplete oxygen burning---overabundances of Si, S, Ca as well as O; \cite{ta70}---and the velocity signature of plumes from incomplete overturn. 
 This picture was extended with x-ray data \citep{casAxray_becker}, and
 Chandra observations \citep{hughes_casA00}, which showed plumes, and an overabundance of Fe---which was predicted to be formed as radioactive $\rm ^{56}Ni$ from explosive Si-burning \citep{tac67,wac73}.
Analysis by \cite{dopita87} indicated overabundaces of Ar---explosive O burning, and Ne---predicted from explosive C burning; \cite{wda69,wzw78}.
The radiactive nucleus $\rm ^{44}Ti$ is a key link in the chain of reactions leading to the iron peak in an explosion \citep{tac67}, and was predicted to have an abundance that might be detected, as observations by  
NuStar \citep{nustar,ti44} have now confirmed.
Hubble Space Telescope (HST) observations \citep{fesen2016} have refined and extended our understanding of the velocity-composition structure. Reliably connecting 3D simulations of the explosion to this observed structure is an ongoing challenge \citep{pay08}, and one which requires self--consistent 3D models of both the progenitor (see \S\ref{s-progen}, \S\ref{s-predyn}) and the collapse--explosion process (\S\ref{s-collapse}, \S\ref{s-thermonuclear}).


\subsection{Mixing boundaries and core helium burning}\label{Shebnd}
The CoRot and Kepler satellites have produced asteroseismological data which strain the present generation of stellar models. See \cite{aerts} for an overview, and  \cite{jt2015} and \cite{tom15,tom16} for some examples.
This is particularly acute for the stage of core helium burning. Asteroseismology indicates that the real stellar cores during He-burning are significantly larger than those predicted, and that adjustment of parameters which were satisfactorily set during hydrogen burning, is no longer adequate. Large adjustments of these parameters 
(``overshoot") within the accepted algorithmic framework produces implausible results, but still does not resolve the discrepancy with asteroseismology. It apprears that a better understanding of the boundary physics during He burning is required. 

The 3D simulations indicate that the Schwarzschild criterion as normally used will underestimate the size of the mixing region, and this error may be exacerbated by the opacity behavior of the He--burning ashes. H-burning produces He, which reduces the opacity; He--burning produces C and O, which increase the opacity. An increased opacity tends to increase convective mixing and core size, so that there may be positive feedback on mixing during core He burning. 

\section{Conclusion}\label{Sconclusion}


\subsection{Weather and climate}
While 3D simulations of the evolution of supernova progenitors is becoming feasible, this is not true for earlier stages. Core convection during main sequence hydrogen burning involves many ($\geq 10^9$) turnover times to complete the exhaustion of fuel. The long--term evolution of a non-linear quasi--steady--state system is a difficult problem; it involves the challenge of extrapolating the {\em weather} to determine the {\em climate}.  The steady-state approximation \citep{am11turb} discussed in \S\ref{Sresults} is plausible but of unproven validity for stars.

\subsection{Effective Reynolds number}
Finite grid resolution requires a loss of information below the grid scale, so that numerical simulations have an {\em effective} Reynolds number which is determined by this ``numerical viscosity''. The real Reynolds number in stars (Re $\sim 10^9$ or more) is much larger than the effective Reynolds number in the best--resolved simulations (Re $\sim 10^3$), so that we are forced to assume that it is valid to extrapolate the computed behavior to the stellar case \citep{amv2014}. However, terrestrial experiments at the largest Reynolds numbers may not support this extrapolation \citep{dim01}. At present we must extrapolate, but should also treat the results with due caution. 

The Reynolds number is Re$\sim u \ell/\nu$, where $u$ is a characteristic flow velocity, $\ell$ is a characteristic length, and $\nu$ is the kinematic viscosity. If we define a numerical viscosity by $\nu_{num} \sim u \Delta r$, where the new length $\Delta r$ is a typical zone size, we may define the corresponding effective Reynolds number as  Re$_{eff} \sim \ell /\Delta r$, or the inverse of the linear resolution. Laboratory fluids exhibit a transition from laminar to turbulent flow at a critical value of Re; it varies with problem, but is $\sim 100$ or so.
The numerical resolution study mimics this behavior (Table~\ref{table1}), with increasing resolution allowing more of the turbulent cascade to be manifest \citep{viallet2015}.
Simulations of low resolution may be badly misleading because of their high numerical viscosity, which tends toward laminar flow, suppressing turbulence. 



\subsection{Carbon, neon and silicon burning}
 Stellar evolutionary models of H--burning and He--burning may be tested by conventional astronomical observations; for example, by their positions in the Herzsprung--Russell (HR) diagram, by observations of eclipsing binaries, and by asteroseismology (g-modes and mixed modes; \S\ref{Shebnd}). However the stages of C, Ne, O, and Si burning are short--lived because of neutrino--cooling, and must be tested in other ways. Because of the different P\`eclet numbers, calibrations from earlier photon--cooled stages will be invalid.
Fortunately direct numerical simulations of these stages are feasible. We have results from O burning
\citep{ma07b,viallet2013,321D}, but are just beginning to simulate C burning \citep{andrea2015}, Neon burning, and Si burning \citep{couch2015}. Because each stage is different, each needs to be studied with realistic turbulent flow. Preliminary indications are that C, Ne and O burning have a family resemblence, but Si burning and the ensuing neutronization by electron capture are clearly somewhat different; see \S\ref{s-collapse} and \cite{couch2015}. 

\subsection{Entropy structure of atmospheres}
In a recent summary of the character of 3D simulations of stellar atmospheres, \cite{magic2016} has shown that they may be characterized by two parameters, the entropy at depth, and the gradient of entropy just below the surface. This important result may be connected to the conventional parameterization of stellar atmospheres by gravity $g$ and effective temperature $T_{\rm eff}$. We note that $L/4 \pi r^2 = F = \sigma T_{\rm eff}^4$ and
$g = -{1 \over \rho} dP/dr $. The flux at the photosphere is a measure of the radiation entropy,
$F = [{cg \beta \over 4 {\cal R} Y \kappa}] S_{rad}$, where the term in square brackets is slowly varying \citep{wda96}. 
The gravititational acceleration $g$ is a measure of the gradient of radiation entropy through the pressure gradient in hydrostatic equilibrium. While not yet exact, these approximations do catch the trend of the simulations \citep{magic2016}, and may offer a significant improvement over fits to the flawed MLT, which lacks turbulent kinetic energy flux and realistic boundary physics.

%
%
%
%


\subsection{Summary}
Numerical simulations in 3D, now appearing for many different astrophysical problems, have become sufficiently mature to show new aspects of turbulent flow in stable and explosive stages of stellar evolution, as well as  some errors in the conventional 1D theory. 
The issues of resolution, and behavior over long time scales, are basic challenges for a number of the problems discussed. With increasing computational power and better algorithms, the simulations are becoming truly turbulent, a significant step closer to reality. Some theoretical analysis of the simulations has been successful, leading to insight beyond the numbers.

\begin{acknowledgements}

This work was supported in part by the Theoretical Program in Steward Observatory.
This work used the Extreme Science and Engineering Discovery Environment (XSEDE), which is supported by National Science Foundation grant number OCI-1053575, and made use of ORNL/Kraken and  TACC/Stampede.
This work was supported in part by resources provided by the Pawsey Supercomputing Centre with funding from the Australian Government and the Government of Western Australia, and 
through the National Computational Infrastructure under the National Computational Merit Allocation Scheme.
\end{acknowledgements}

\end{document}